\documentclass[prl,twocolumn,showpacs]{revtex4}
\usepackage{amsmath}
\usepackage{graphicx} 
\usepackage{dcolumn }
\usepackage{mathptmx}  

\begin{document}

\title{The chiral phase transition in charge ordered {\it 1T}-TiSe$_2$}

\author{John-Paul Castellan$^1$}
\author{Stephan Rosenkranz$^1$}
\author{Ray Osborn$^1$}
\author{Qing'an Li$^1$}
\author{K.E. Gray$^1$}
\author{X. Luo$^1$}
\author{U. Welp$^1$}
\author{Goran Karapetrov$^{1,2}$}
\author{J.P.C. Ruff$^{3,4}$}
\author{Jasper  van Wezel$^{1,5}$}
\email{Jasper.vanWezel@bristol.ac.uk}
\affiliation{
$\phantom{\text{Materials Science Division, Argonne National Laboratory, Argonne, IL 60439, USA.}}$\\
$^1$ Materials Science Division, Argonne National Laboratory, Argonne, IL 60439, USA.\\
$^2$ Physics Department, Drexel University, Philadelphia, PA 19104, USA.\\
$^3$ Advanced Photon Source, Argonne National Laboratory, Argonne, IL 60439, USA.\\
$^4$ CHESS, Cornell University, Ithaca, NY 14853, USA.\\
$^5$ H.H. Wills Physics Laboratory, University of Bristol, Bristol BS8 1TL, U.K.
}

\begin{abstract}
It was recently discovered that the low temperature, charge ordered phase of {\it 1T}-TiSe$_2$ has a chiral character. This unexpected chirality in a system described by a scalar order parameter could be explained in a model where the emergence of relative phase shifts between three charge density wave components breaks the inversion symmetry of the lattice. Here, we present experimental evidence for the sequence of phase transitions predicted by that theory, going from disorder to non-chiral and finally to chiral charge order. Employing X-ray diffraction, specific heat, and electrical transport measurements, we find that a novel phase transition occurs $\sim 7$~K below the main charge ordering transition in TiSe$_2$, in agreement with the predicted hierarchy of charge ordered phases.
\end{abstract}


\maketitle

{\it Introduction.}--The modulation of electronic density which emerges in charge ordered materials reduces the translational symmetry of the underlying lattice. Additional symmetries may be broken through the coupling to atomic displacements, and these are often key in understanding the material properties of charge ordered systems. The broken rotational symmetry in {\it 2H}-TaSe$_2$, for example, yields a reentrant phase transition under pressure \cite{Littlewood82}, while the broken inversion symmetry in rare-earth nickelates {\it R}NiO$_3$, renders them multiferroic \cite{VanDenBrink09}.

Recently, it was discovered that the breakdown of inversion symmetry in the charge ordered phase of {\it 1T}-TiSe$_2$ leads to the presence of a chiral structure at low temperatures \cite{Ishioka10,VanWezel10,VanWezel11,VanWezel12:2,Ishioka11,VanWezel12:3,Iavarone12}. In this phase, a helical charge density distribution arises from a rotation of the dominant charge density wave component as one progresses through consecutive atomic layers. While helical phases are common among spin-density waves, with vectorial order parameters, {\it 1T}-TiSe$_2$ is one of only very few materials so far in which a chiral charge ordered phase, with a scalar order parameter, has been suggested to exist \cite{Guillamon11,VanWezel12}. A mechanism for the formation of this scalar chirality in TiSe$_2$ was recently proposed, in which the chiral phase is interpreted to be simultaneously charge and orbital ordered \cite{VanWezel11}.

The scanning tunneling microscopy experiments in which the chirality of TiSe$_2$ was initially revealed, were performed well below the onset temperature of charge order in this material \cite{Ishioka10}. There is currently no experimental information on the nature of the transition between the chiral and the normal state.  Here, we present experimental evidence for a sequence of two transitions, the well-known onset of charge order at $\sim 190$~K and a novel transition at $\sim 183$~K.  The temperature dependence of X-ray superlattice reflections unnoticed in previous studies, combined with an analysis of their structure factors, and the temperature dependences of the specific heat and resistance anisotropy, strongly suggests a scenario in which the transition at $\sim 183$~K indicates the emergence of chiral charge order out of the non-chiral charge density wave state.  The existence of this hierarchy of transitions, as well as the observed experimental signatures, are in agreement with recent theoretical predictions \cite{VanWezel11,VanWezel12:2}.

\begin{figure}
\begin{center}
\includegraphics[width=0.97\columnwidth]{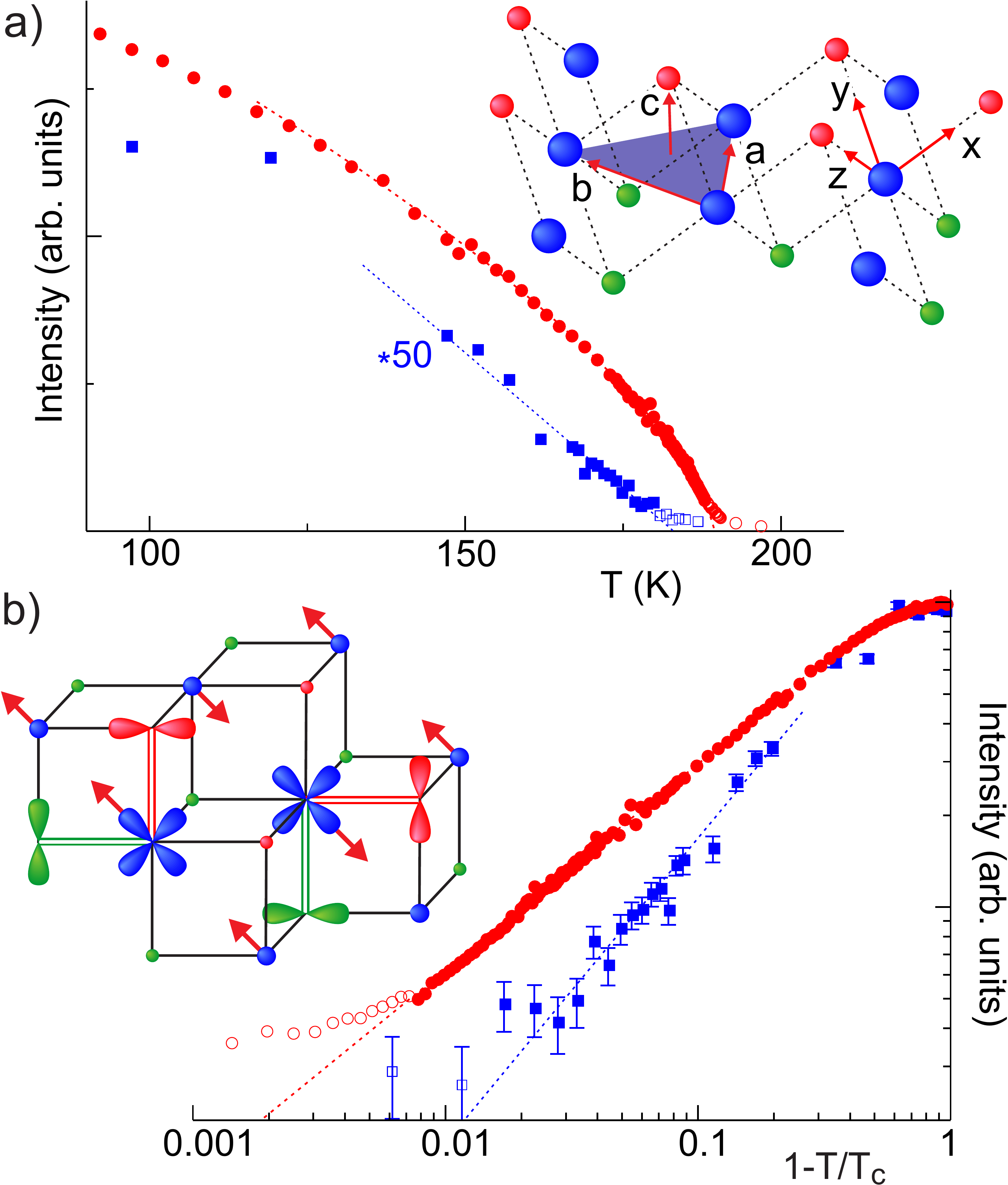}
\end{center}
\caption{(Color online) {\bf a)} The integrated intensities of the superlattice reflections at  [$\frac{3}{2}$,$\frac{3}{2}$,$\frac{1}{2}$] (upper, red curve) and [$\frac{5}{2}$,$1$,$0$] (lower, blue curve; amplified by a factor of $50$) \cite{footnote}. {\bf inset)}  Schematic depiction of a single layer of TiSe$_2$. The plane of Ti atoms (large and blue) is indicated along with the crystallographic axes $a$, $b$ and $c$. The orbital basis instead uses $x$, $y$ and $z$ axes, which connect the Ti and Se atoms (smaller, red and green). {\bf b)} The same data on a log-log scale, as a function of $1-T/T_{\text{CDW}}$ (red data) or $1-T/T_{\text{Chiral}}$ (blue), along with power law fits. The error bars for the red curve do not exceed the size of the symbols. In both panels, the open symbols form a fluctuation tail, extending beyond $T_C$. {\bf inset)} Depiction of one of the individual charge density wave components.}
\label{Xray}
\end{figure}

{\it Chiral Charge Order.}---{\it 1T}-TiSe$_2$ is a quasi-two dimensional material, in which hexagonal layers of Ti are sandwiched between similar layers of Se, and individual sandwiches are separated from each other by a large van der Waals gap (see Fig.~\ref{Xray}a). It is observed to undergo a phase transition into a commensurate charge density wave state, accompanied by periodic lattice distortions. The mechanism underlying the transition is heavily debated \cite{Rossnagel11,Cercellier07,VanWezel:EPL,VanWezel10:2,Weber11}. Regardless of what drives its formation, however, it is clear that the charge density modulation consists of three components, each involving a charge transfer process between one particular Ti-$3d$ orbital and two Se-$4p$ orbitals. One such component is depicted in Fig.~\ref{Xray}b. A non-zero phase shift of this component corresponds to more charge being transferred along the red upper than along the green lower bonds, and a corresponding translation of the displacement wave.

If the three components are superposed without any relative phase differences, the 2x2x2 non-chiral charge density wave state originally proposed for TiSe$_2$ is produced \cite{DiSalvo76}. Superposing three components with different phase shifts result in a chiral 2x2x2 lattice distortion \cite{VanWezel11}. Whether the chiral or the non-chiral state is energetically favorable, is determined by minimizing the Landau free energy. Writing the order parameter as the sum of three complex parameters $\psi_j = \psi_0 e^{i \vec{q}_j \cdot \vec{x} + \varphi_j}$ with equal amplitudes, the free energy may be written as:
\begin{align}
F = & \psi_0^2 [ a_0 (T/T_{\text{CDW}}-1) + 2 a_1 \sum_j \cos^2(\varphi_j) ] \notag \\
+ & \psi_0^4 [ b_0 + 2 b_1 \sum_j \cos^2(\varphi_j - \varphi_{j+1}) ].
\label{F}
\end{align}
Here $a_0$ and $b_0$ represent the combined effects of Coulomb interaction and the competition between charge density wave components, while the terms $a_1$ and $b_1$ are the leading order Umklapp terms which signify the coupling between the electronic order parameter and the atomic lattice \cite{VanWezel11}. 

Minimizing the free energy with respect to the phase variables yields two possible solutions. Coming from high temperatures, the non-chiral state with $\varphi_1=\varphi_2=\varphi_3$ is first realized at $T=T_{\text{CDW}}$, when the amplitude $\psi_0$ becomes non-zero. At the lower temperature $T_{\text{Chiral}} = T_{\text{CDW}} ( 1 - \frac{2 a_1}{3 a_0} [6 + \frac{b_0}{b_1} ] )$, the differences between the three phase variables $\varphi_j$ become non-zero, and the chiral charge order sets in at a second-order phase transition. Since the contribution from Umklapp effects is generally weaker than that of the direct interactions, $1 - T_{\text{Chiral}} / T_{\text{CDW}} \propto \frac{a_1}{a_0}$ may be expected to be small \cite{VanWezel11}.

{\it X-ray diffraction.}---Single crystals of TiSe$_2$ were prepared using the iodine vapor transport method \cite{Oglesby94}. A sample of $3$x$3$x$0.05$~mm$^3$ was mounted on the cold finger of a closed cycle displex and aligned on a Huber six-circle diffractometer.  X-ray measurements utilizing area detectors on the sector 6-ID-D high energy station of the APS were performed in transmission geometry, using an incident photon energy of $80$~keV. Detailed order parameter measurements of the charge ordered phase were made using the superlattice reflection [$\frac{3}{2}$,$\frac{3}{2}$,$\frac{1}{2}$], which is indicative of a doubling of the unit cell in all crystallographic directions. The three-dimensional integrated intensity of these scans is presented in Fig.~\ref{Xray}a. Notice that the transition temperature $T_{\text{CDW}}=190$~K is slightly lower than the optimal transition temperature reported for TiSe$_2$, indicating a slight deviation from stoichiometry~\cite{Taguchi81}.

We then used area detectors to map out large volumes of reciprocal space to search for evidence of the emerging chiral order below $T_{\text{CDW}}$.  We discovered extra peaks  in the low temperature maps at wave vectors of the form [$H+\frac{1}{2}$,$K$,$0$], with intensities $\sim 50$ times weaker than the primary charge density wave reflections. We measured one of these peaks, located at the [$\frac{5}{2}$,$1$,$0$] superlattice reflection, as a function of temperature.

Notice that although this reflection is not forbidden by symmetry in the non-chiral space group, it has a low intensity due to its small structure factor $S({\bf q})$. Up to an element-specific form factor, the contribution of each species of atom to the structure factor is determined by the Fourier transform of their positions in the crystal lattice. Using the predicted displacements for each of the charge density wave components (as shown in Fig.~\ref{Xray}b), it becomes clear that, for any given amplitude of the distortion, the contributions to the structure factor for the peak at [$\frac{5}{2}$,$1$,$0$] are significantly greater in the chiral configuration than in the corresponding non-chiral charge ordered state: $S^{Ti}_{Chiral}([\frac{5}{2},1,0]) / S^{Ti}_{CDW}([\frac{5}{2},1,0]) \simeq 1.86$ while $S^{Se}_{Chiral}([\frac{5}{2},1,0]) / S^{Se}_{CDW}([\frac{5}{2},1,0]) \simeq 64$, which implies a pronounced enhancement of the diffraction intensity at $[\frac{5}{2},1,0]$ as the chirality sets in, as seen in Fig.~\ref{Xray}a. 

The evolutions of the two peaks under decreasing tem- perature depicted in Fig.~\ref{Xray}a show that the intensity at [$\frac{3}{2}$,$\frac{3}{2}$,$\frac{1}{2}$] first becomes non-zero at $T_{\text{CDW}}=190$~K. The intensity of this peak is expected to be proportional to the squared amplitude of the charge density wave order parameter. The intensity of the reflection at [$\frac{5}{2}$,$1$,$0$] on the other hand, remains zero below the onset temperature $T_{\text{CDW}}$, and begins to gain intensity only at the lower temperature $T_{\text{Chiral}}=183$~K, which we identify as a second phase transition.

The evolution of the two peak intensities with decreasing temperature can be clearly seen in Fig.~\ref{Xray}b to be significantly different. The intensity of the main charge density wave peak close to $T_{\text{CDW}}$ is best fitted by the power law $I=I_0 (1-T/T_{\text{CDW}})^{2 \beta}$, with $2 \beta \simeq 0.68$. The second peak on the other hand, is a second harmonic since it necessarily involves two components of the charge density wave. Its intensity therefore scales with the fourth power of $\beta$, and is best described by  a power law with $4 \beta \simeq  1.0$. The difference between the best fits for these critical exponents is suggestive of the onset of a different type of order at $T_{\text{Chiral}}$, or $\sim 7$ K below the initial onset of charge order at $T_{\text{CDW}}$. Signatures of this second phase transition are also seen in specific heat and resistivity measurements, as described below.

\begin{figure}
\begin{center}
\includegraphics[width=\columnwidth]{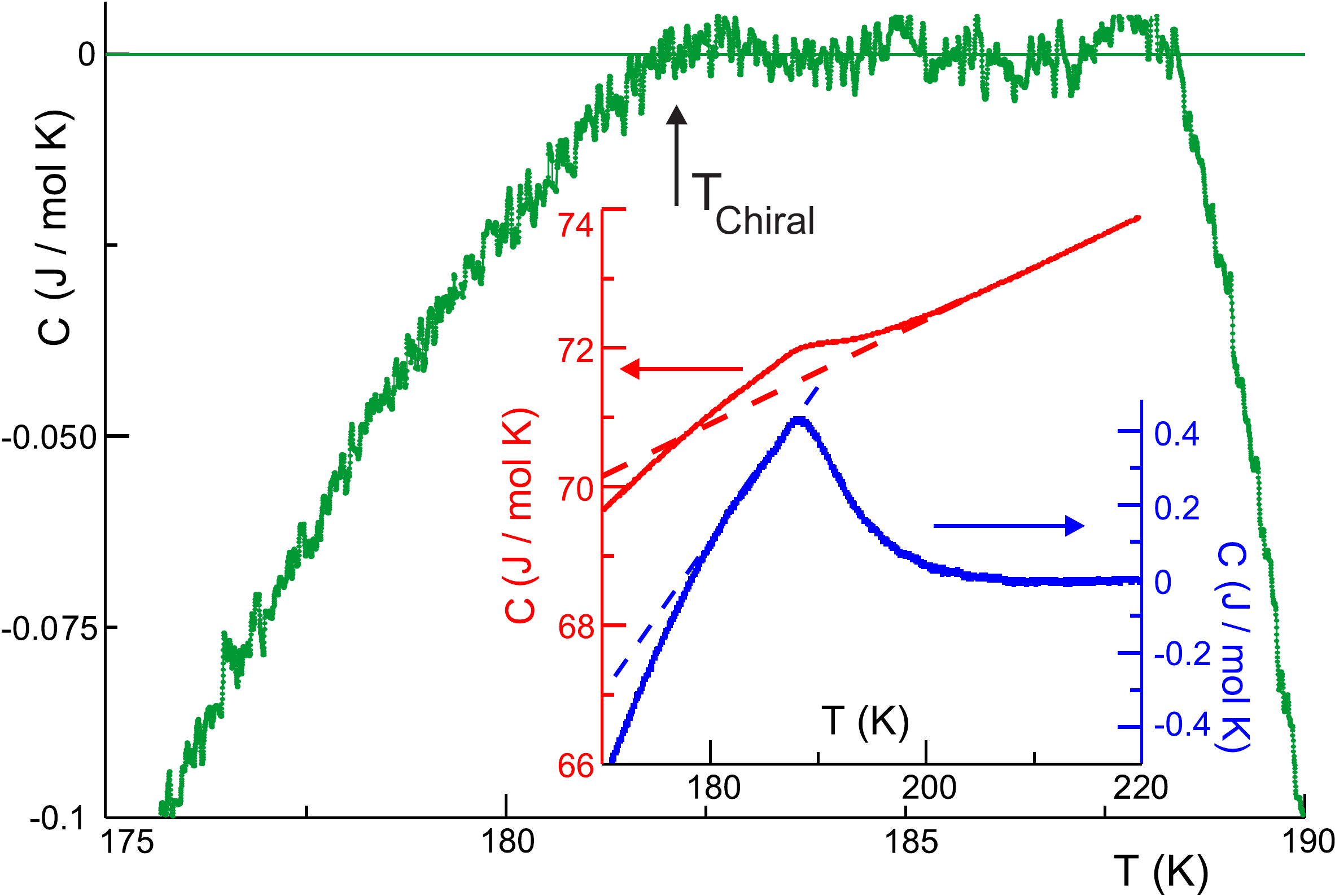}
\end{center}
\caption{(Color online) The specific heat as a function of temperature. {\bf inset)} Bare specific heat before (red, upper curve) and after (blue lower curve) subtracting the high temperature background (red dashed line). The blue dashed line emphasizes the linear temperature dependence below the initial onset of charge order at $T_{\text{CDW}}$, as well as the break in slope which occurs at the lower temperature $T_{\text{Chiral}}$. This is shown more clearly in the main panel, which displays the specific heat after subtracting the linear fit below $T_{\text{CDW}}$.}
\label{heat}
\end{figure}

{\it Specific Heat.}---
The temperature dependence of the specific heat of a $165$x$165$x$7$~$\mu$m$^3$  TiSe$_2$ crystal cut from the sample used in the X-ray diffraction and electric transport experiments, is shown in Fig.~\ref{heat}.  We performed calorimetric measurements using a membrane-based steady-state ac-micro-calorimeter~\cite{Welp1,Welp2,Tagliati}, with a thermocouple composed of Au-$1.7\%$Co and Cu films deposited onto a $150$~nm thick Si$_3$N$_4$-membrane as a thermometer. This technique enables precise measurements of changes in specific heat. The absolute accuracy of our data was checked against measurements on gold samples of similar size as the samples studied here.  The upper, red curve in the inset of Fig.~\ref{heat} displays the bare specific heat data.  An anomaly around $\sim 190$~K indicating the charge density wave transition is clearly seen.  Plotting the data with respect to a linear extrapolation of the high-temperature background (the red dashed line) yields a detailed presentation of the specific heat anomaly (lower, blue curve in the inset).  The height of the anomaly of $\sim 0.5$~J/mol~K and the extended high-temperature tail are similar to previous reports~\cite{Welp3}. This tail induces some uncertainty into the determination of the transition temperature $T_{\text{CDW}}$. The inflection point yields $T_{\text{CDW}} \sim 191$~K whereas the specific heat peak is at $\sim 189$~K.  Remarkably, the specific heat is linear in temperature below the peak position, as highlighted by the blue dashed line, followed by a kink near $\sim182$~K.  This kink is clearly seen in the main panel, which presents the data plotted with respect to the blue dashed line.

The high-temperature tail of the charge density wave transition indicates the presence of fluctuations or a distribution of transition temperatures.  At present, these effects cannot be separated from the mean-field anomaly of the specific heat, and a quantitative analysis based on the free energy expression in Eq.~\eqref{F} is not possible. However, the break in the slope of the specific heat signals a change in the thermodynamic state of the sample, which we identify with the transition into the chiral phase. This is supported by the mean-field behavior predictions of Eq.~\eqref{F}. Taking the second derivative with respect to temperature for the lowest energy solutions yields linear temperature dependences for both $T_{\text{Chiral}} < T < T_{\text{CDW}}$ and for $T < T_{\text{Chiral}}$, but with different slopes:
\begin{align}
C_{V} = \left\{ \begin{array}{lcr} \frac{T}{T_{\text{CDW}}^2} \frac{a_0^2}{2 b_0+12 b_1}, & \ & T_{\text{Chiral}} < T < T_{\text{CDW}} \\ \frac{T}{T_{\text{CDW}}^2} \frac{a_0^2}{2 b_0+3 b_1}, & & T < T_{\text{Chiral}} \end{array} \right.
\label{C}
\end{align}
We anticipate that this break in slope is apparent in the data even when fluctuations and inhomogeneous broadening are superimposed. The additional step-like discontinuity expected at the chiral transition is not clearly resolved in the data, either because our present experimental resolution is insufficient, or because it is masked by broadening due to the fluctuations surrounding the transition, or due to an increased sample inhomogeneity resulting from the formation of domain walls in the chiral charge ordered state. We note that the high-temperature tails above $T_{\text{CDW}}$ as well as $T_{\text{Chiral}}$ are also apparent in the X-ray diffraction data, and that the two transition temperatures identified in the specific heat data closely match those suggested by the X-ray measurements.
\begin{figure*}
\begin{center}
\includegraphics[width=\textwidth]{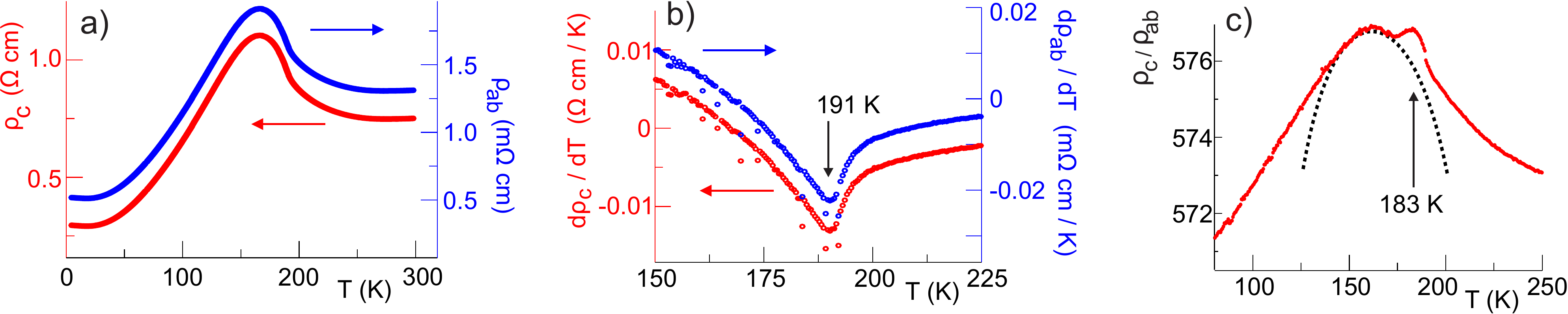}
\end{center}
\caption{(Color online) {\bf a)} The principal components of resistivity as a function of temperature. Notice that the scales for the two curves differ by a factor of $\sim 575$. {\bf b)} The derivatives of both curves, displaying sharp minima at $T_{\text{CDW}}$, the initial onset  temperature of (non-chiral) charge order. {\bf c)} The resistivity anisotropy $\rho_c / \rho_{ab}$ as a function of temperature. The peak at $183$~K, defining the second transition temperature $T_{\text{Chiral}}$, can be clearly identified. The broad feature centered at $161$~K follows from the presence of broad maxima in $\rho_c(T)$ and $\rho_{ab}(T)$, and does not signify an additional transition. It can be modeled by fitting the broad peaks in panel {\bf a} to parabolic functions, resulting in the dotted line.}
\label{resistivity}
\end{figure*}

{\it Electrical Transport.}---In a layered material like TiSe$_2$, traditional four-terminal methods to determine the resistivity along the $c$-axis, $\rho_c$, and in the $ab$-plane, $\rho_{ab}$, may be unreliable for single crystals, especially if the conductance anisotropy is large.  This problem is overcome by a six-terminal method that uses a single crystal in a rectangular shape \cite{Busch92}.  We employed four parallel Au stripes for electrical contacts, that were sputtered onto each $c$-axis-normal surface, perpendicular to the longest in-plane dimension of the crystal.  The current was injected through the outermost contacts of one surface and voltages were measured across the innermost contacts of each surface.  The Laplace equation was then solved and inverted to find $\rho_{ab}$ and $\rho_c$ \cite{Busch92,Li99}. This method also allows a test of sample homogeneity by permuting the electrodes used for current and voltage \cite{Li07}.

Fig.~\ref{resistivity}a  shows the principle components of the resistivity for the same TiSe$_2$ crystal used in the X-ray diffraction studies. The maximum variation upon permuting the electrodes is less than $5$~\%, implying a high degree of homogeneity. The temperature profiles of $\rho_{ab}$ and $\rho_c$ are also almost identical, although their absolute values differ by an anisotropy factor of $\sim 575$.  Each resistivity component exhibits a maximum close to $168$~K, which has previously been suggested to arise from an initial decrease in the density of available carriers, caused by the opening of a gap in the charge ordered phase, which is overtaken at lower temperatures by both the decrease of scattering channels due to the developing order, and an increase in density of states due to the downward shift of the conduction band minimum below $T_{\text{CDW}}$ \cite{Monney10}. The position of the maximum thus does not coincide with any charge ordering transition. Instead, Di Salvo et al. found that a sharp minimum in $d\rho/dT$ matches the onset of the charge ordered superlattice as determined by neutron scattering \cite{DiSalvo76}. This is confirmed by our data in Fig.~\ref{resistivity}b, which show $d\rho_c/dT$ and $d\rho_{ab}/dT$ to be virtually coincident after scaling by the anisotropy, with a minimum at $191$~K that closely matches the onset of charge order as determined by X-ray diffraction (see Fig.~\ref{Xray}a).

While $\rho(T)$ in Fig.~\ref{resistivity}a varies by more than a factor of three, the temperature variation of the resistivity anisotropy $\rho_c / \rho_{ab}$ shown in Fig.~\ref{resistivity}c, is significantly smaller at only $\sim 1.5$~\%.  This observation agrees with the suggestion that the overall temperature dependence of $\rho(T)$ is determined primarily by the evolution of the carrier density \cite{Monney10}, while the variations in anisotropy are determined by the temperature dependence of the scattering, which evidently plays a much smaller role. The anisotropy exhibits a sharp peak at $183$~K and a relatively broad maximum centered at $161$~K. The latter feature is caused by slight differences in position and height of the maxima in $\rho_c(T)$ and $\rho_{ab}(T)$, and does not signify a qualitative change in physics. It can be modeled by parabolic fits of the maxima in Fig.~\ref{resistivity}a, which results in the anisotropy indicated by the dotted curve in Fig.~\ref{resistivity}c. The sharp feature at $183$~K on the other hand, closely matches the second transition temperature identified in our X-ray diffraction and specific heat experiments. 

The onset of chiral order will be accompanied by the formation of right and left-handed domains. These have been shown to occur in both pristine and Cu-intercalated TiSe$_2$, using (surface sensitive) scanning tunneling microscopy~\cite{Ishioka10, Ishioka11, Iavarone12}, as well as (bulk) reflectivity measurements~\cite{Ishioka10}. The typical sizes of the observed domains are of the order of tens of nanometers~\cite{Iavarone12}. The corresponding presence of domain walls within the $a,b$-plane may be expected to lead to a sharp increase of electronic scattering within the plane, without affecting electrical transport along the $c$ axis, thus causing a decrease of the resistivity anisotropy $\rho_c / \rho_{ab}$ with decreasing temperature below $T_{\text{Chiral}}$. The feature at $183$~K in the anisotropy in Fig.~\ref{resistivity} is thus consistent with the expected increase of $\rho_{ab}$ below $T_{\text{Chiral}}$ resulting from increased scattering off chiral domain walls.

{\it Conclusions.}---We conclude that the low temperature charge ordered state in {\it 1T}-TiSe$_2$ arises in a sequence of two closely separated phase transitions. The first transition, at $T_{\text{CDW}} \simeq 190$~K, is well known to indicate the onset of a 2x2x2 charge ordered state. The second transition, which we identify to occur at $T_{\text{Chiral}} \simeq 183$~K, is characterized by the emergence of previously unobserved X-ray diffraction peaks, a sudden change in slope of the specific heat, and a sharp peak in the resistivity anisotropy. The temperature dependence of the X-ray superlattice reflections, combined with an analysis of their structure factors, and the temperature dependence of the specific heat and resistivity, strongly suggests a scenario in which the  transition at $\sim 183$~K indicates the emergence of chiral charge order out of the non-chiral charge density wave state.  
This sequence of two transitions leading first from the normal state to the non-chiral charge density wave, and only then to the chiral charge ordered state, agrees with the predictions arising from the theoretical model of Ref.~\cite{VanWezel11}, which describes the recently discovered chiral state of {\it 1T}-TiSe$_2$ in terms of combined charge and orbital order.

\subsection{Acknowledgements}
Work at the Advanced Photon Source and Material Science Division of Argonne National Laboratory were supported by the U.S. DOE-BES under Contract No. NE-AC02-06CH11357.


\begin{thebibliography}{26}
\bibitem[{}]{footnote}
\bibinfo{note} {A $3$~K shift in temperature was applied to all raw X-ray diffraction data in order to correct for a temperature gradient between sample and diode, located $\sim 4$~cm apart in the cold head}.
\end{thebibliography}
\end{document}